\input{psfig}
\def\HI{H{\,\small I}}
\def\OIII{[O{\,\small III]}}
\newcommand{\kms}{$\,$km$\,$s$^{-1}$}

\newcommand{\msun}{{$M_\odot$}}
\documentclass[cup6a]{cupbook}

\title[ AGN Feedback in Galaxy Formation]
      {Proceedings of the Workshop held in Vulcano (Messina), Italy, May 18-22, 2008}
\author{V.Antonuccio-Delogu and J. Silk, eds. \TeX-to-type}
\date{\today}

\begin{document}

\pagenumbering{roman}
\maketitle
\tableofcontents
\cleardoublepage
\pagenumbering{arabic}

\chapter[Interaction and gas outflows in radio-loud AGN]{Interaction and gas outflows in radio-loud AGN - Disruptive and constructive effects of radio jets \\Raffaella Morganti\\
Netherlands Institute for Radio Astronomy (NL)\\
and \\
Kapteyn Astronomical Institute, University of Groningen (NL)}

\section{Why radio-loud AGN?}

In recent years, Active Galactic Nuclei (AGN) have become more popular for a wider community. The possibility of using them to produce feedback effects that would help solving some of the questions connected to the hierarchical scenario of galaxy formation and evolution, has made them particularly popular among theorists.  Feedback effects associated with AGN-induced outflows are now routinely incorporated in models of galaxy evolution. Indeed, gas outflows may have a wide range of effects. For example,  clearing up the circum-nuclear regions and halting the growth of the supermassive black-holes (see e.g.  Silk \& Rees 1998) as well as providing the mechanism that can cause the correlations  found between the mass of the central super-massive black-hole and the properties of the host galaxies.  Outflows can also prevent the formation of too many massive galaxies in the early universe and can inject energy and metals into the interstellar and intergalactic medium.  AGN-driven outflows could be a major source of feedback in the overall galaxy formation process.
However, {\sl the characteristics of such feedback are poorly constrained and the exact relevance of gaseous outflows in galaxy evolution still need to be evaluated}.

AGN-driven outflows can have different origin. Here, I will concentrate on the role that the {\sl radio-loud phase of nuclear activity} (and the presence of radio plasma jets) can play in the evolution of a galaxy. 
Radio-loud AGN are preferentially hosted by massive early-type galaxies.  
This fraction of these galaxies are radio loud increases with mass: for the highest masses, the fraction of
galaxies that are radio source is $\sim 25$\%. Considering that radio-loud AGN live for only $10^7$ - $10^8$ yr, the radio source activity must be constantly re-triggered (Best et al. 2005 and ref. therein).  Indeed many examples are known of objects where signatures of a recurrent radio activity are observed. They include famous cases like Centaurus~A and 3C~236  and less famous cases associated with recently restarted radio sources (see e.g.  Stanghellini 2005). Thus, radio activity can be common in the life of {\sl most} (if not all) early-type galaxies and may, therefore, be relevant in their evolution. 

Before proceeding, it is also worth noting that, although the radio-loud sources considered in this paper are at low redshift and hosted by  early-type galaxies, it  does not mean that feedback effects are not relevant. Recent detailed studies of the ISM in nearby early-type galaxies have shown how these systems can be complex and that {\sl the assumption that these systems are "red and dead" and without an interesting and rich ISM does not hold for many of them}.
Neutral hydrogen, molecular and ionised gas (see Morganti et al. 2006, Combes et al. 2007 and Sarzi et al. 2006 respectively) are often found (sometimes in large quantities) in these galaxies. This is also the case around and in the centre of, at least some, radio sources. The possible connection between the presence of gas and the presence (and type) of the radio source is also matter of investigation (see e.g. Emonts et al. 2006). Relevant in the context of feedback and outflows is the fact that  compact/young radio galaxies are  more often detected in \HI\ (both in emission and in absorption)  and they show more extreme kinematics of the ionised gas (see e.g. Holt et al. 2008 and refs therein) compared to extended/classical radio sources. 

Here, I will concentrate on two aspects in which radio activity could be important. The first is exploring whether relativistic plasma jet  associated with radio-loud galaxies  could provide an effective way to produce gas outflows with characteristics that can be relevant in the evolution of the host galaxy. The second, is to investigate whether they can provide a mechanism for the triggering of  star formation.

The origin of nuclear activity is often explained as connected to merger and/or interaction processes.  These processes  can  bring gas into the system, but they can also trigger radial motions that could lead to increased fuel rate.   In about 30\% of powerful radio galaxies we know, from the analysis of the stellar population, that a gas-rich merger must have occurred in the recent history of the host galaxy. 
Although these are the most extreme cases, one can expect that at least in  the initial phase of activity, the super massive BH will often be surrounded by a rich gaseous medium (indeed observed in the case of young radio sources as mentioned above). 
The relativistic jets associated with the radio-loud phase of activity  can provide the mechanism to directly coupling the AGN to its environment and produce gaseous outflows. These outflows can be, therefore, also relevant for the orientation-independent obscuration,  again providing one of the mechanism (together with radiation pressure and starburst winds)  to expel  the gas that obscure the AGN in the initial phase If this is the case, the outflows would be particularly important for young AGN. Conversely,  the ISM may also have an influence in the evolution of the radio jet: the interaction may cause a (temporarily) disruption of the flow. 
Finally, jet-induced star formation has been suggested to be important for high-$z$ objects and be one of the possible causes of alignment effect.  This effect is not so commonly seen at low-z although there are a few cases known that have been studied in detail. 

In summary, the main aim of the projects described below is to understand what are the main sources of feedback and outflows in the nearby objects and extrapolate this to the high-$z$ Universe where usually the objects cannot be studied in such a details.

\begin{figure}
\centerline{\psfig{figure=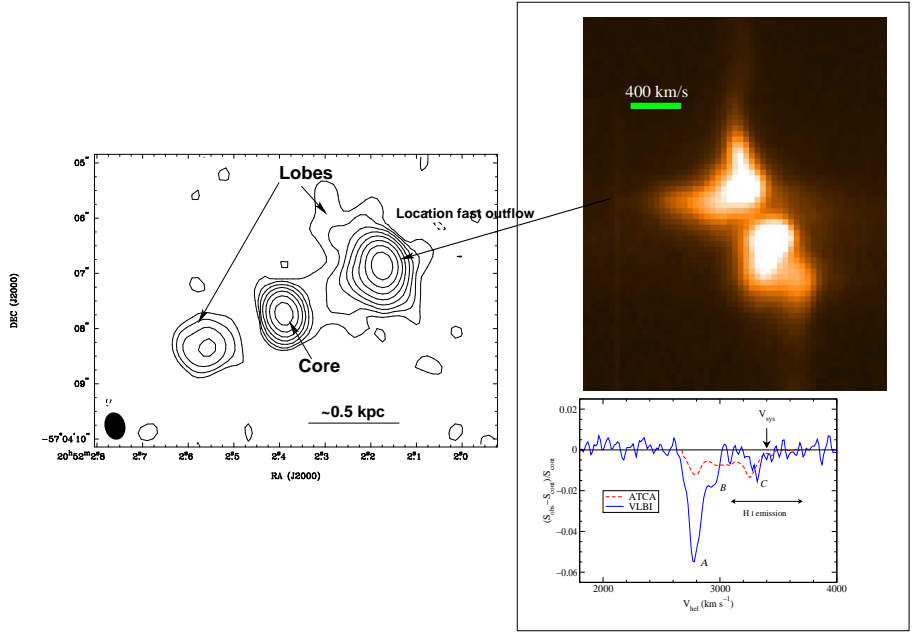,width=12cm,angle=0}
}
    \caption{{\sl Left:} Radio continuum image of IC~5063; {\sl Right:} Comparison between the width of the \HI\ absorption (bottom plot) and that of the ionised gas (top; from the {\OIII}5007\AA). The
  first order similarity between the amplitude of the blueshifted
  component is clearly seen.}
    \label{fig1}
 \end{figure}

\section{The nuclear regions probed by the \HI\ and ionised gas}

As mentioned above, gas is an important ingredient in the regions surrounding an active nucleus.  Thus, we have used the gas (21 cm-\HI\ and ionised gas) to trace the effect of the radio plasma passing into the ISM of a radio source. Here we will summarise some of the best studied cases and the most relevant results.

\subsection{Fast \HI\ outflows: our best studied case - IC~5063}

One of our best studied objects is the radio-loud Seyfert 2 galaxy IC~5063. Recent radio data (see  Morganti et al.  2007 and ref. therein) have confirmed the triple structure of the source with a central, unresolved flat-spectrum core and two resolved radio lobes with steep spectral index (see Fig.~\ref{fig1} - left). This implies that the previously detected fast outflow of neutral gas is occurring off-nucleus, near the (brighter) radio lobe, i.e. about 0.5 kpc from the core.  The ionised gas shows highly complex kinematics in the region co-spatial with the radio emission.  Broad and blueshifted ($\sim 500$ \kms) emission is observed in the region of the radio lobe, at the same location as  the blueshifted \HI\ absorption. The velocity of the ionised outflow is similar to the one found in \HI\ (see Fig.~\ref{fig1} - right). The first order correspondence between the radio and optical properties suggests that the outflow is driven by the interaction  between the radio jet and the ISM.  

\begin{figure}
\centerline{\psfig{figure=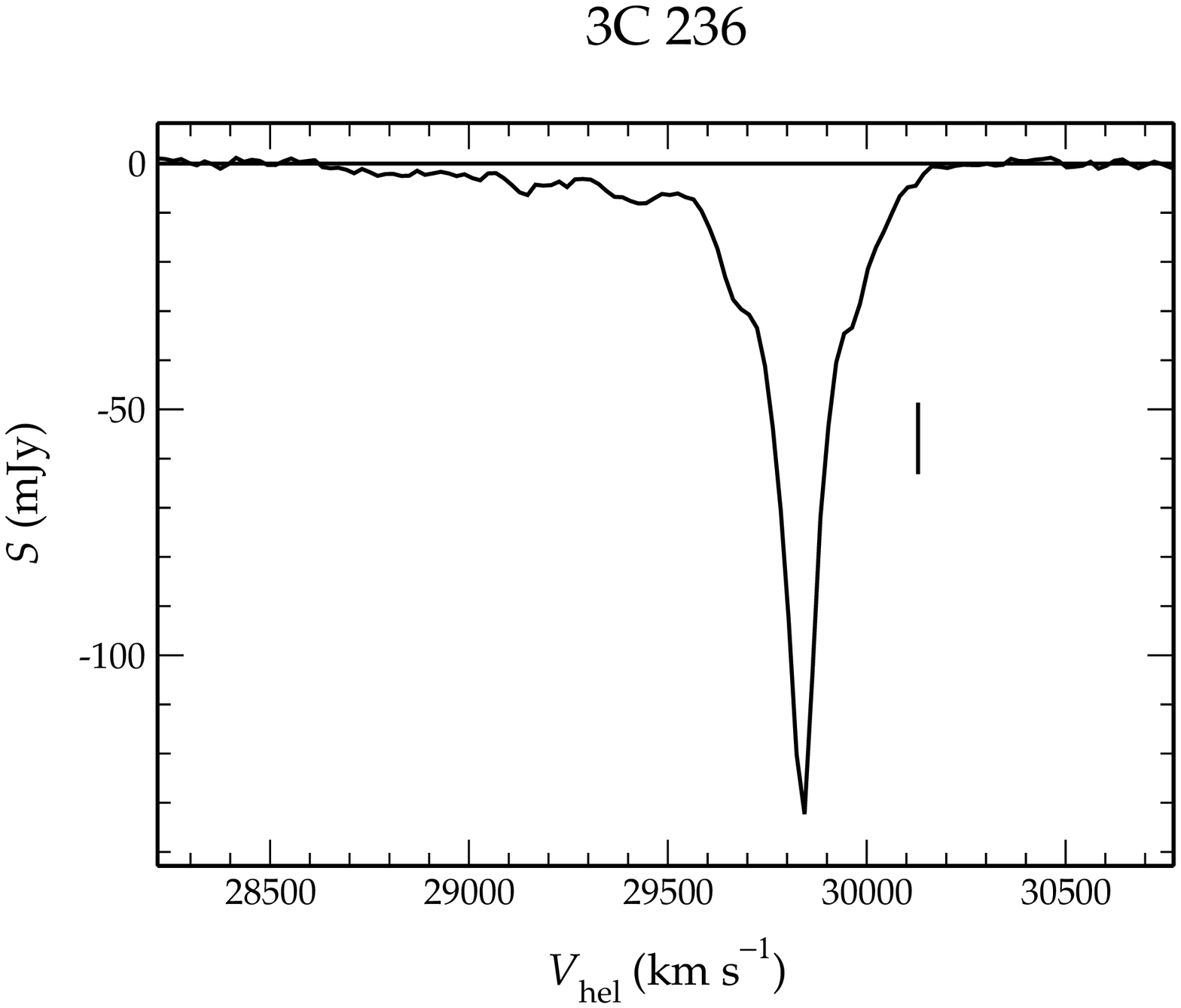,angle=0,width=4.5cm}
\psfig{figure=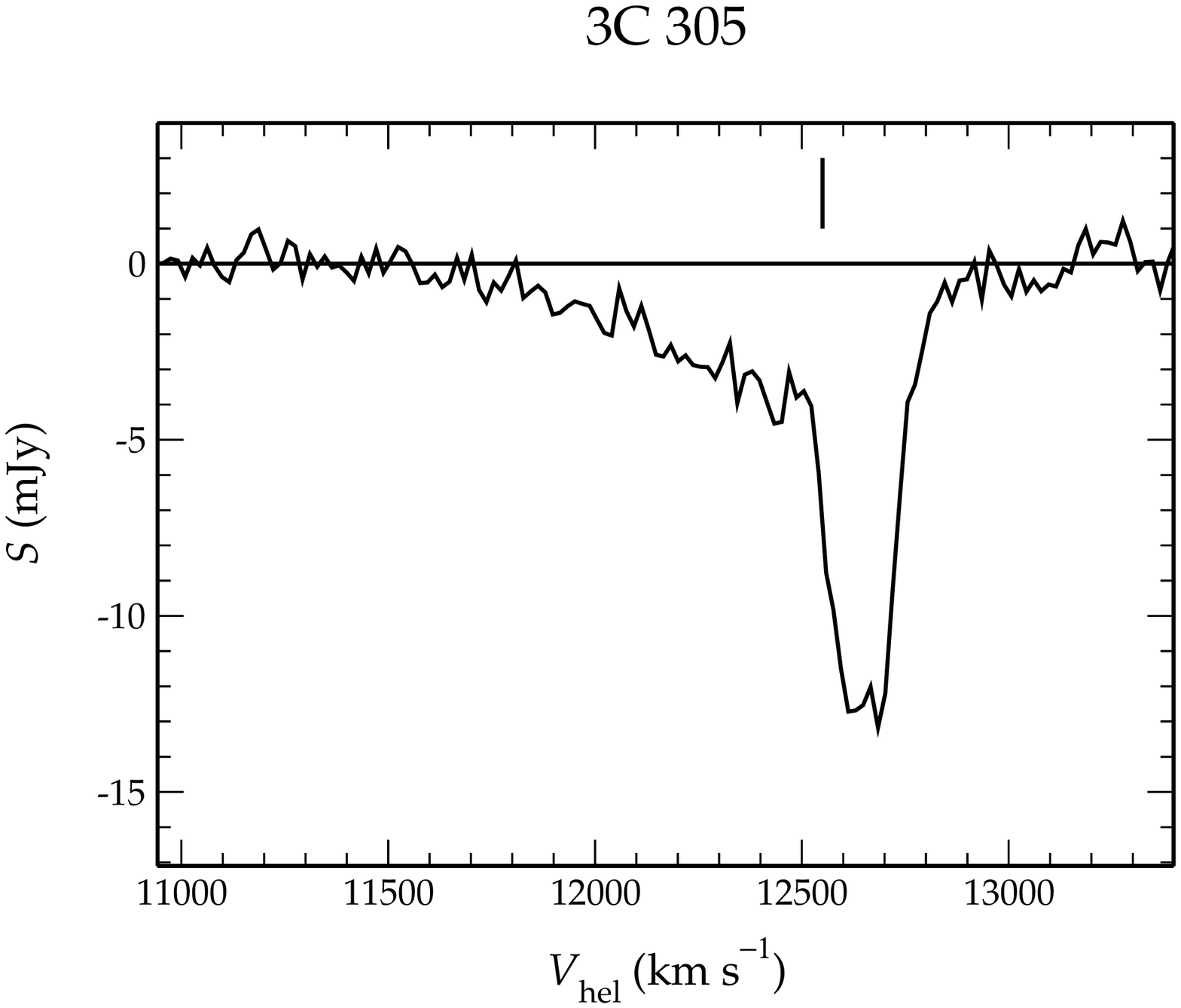,angle=0,width=4.5cm}
\psfig{figure=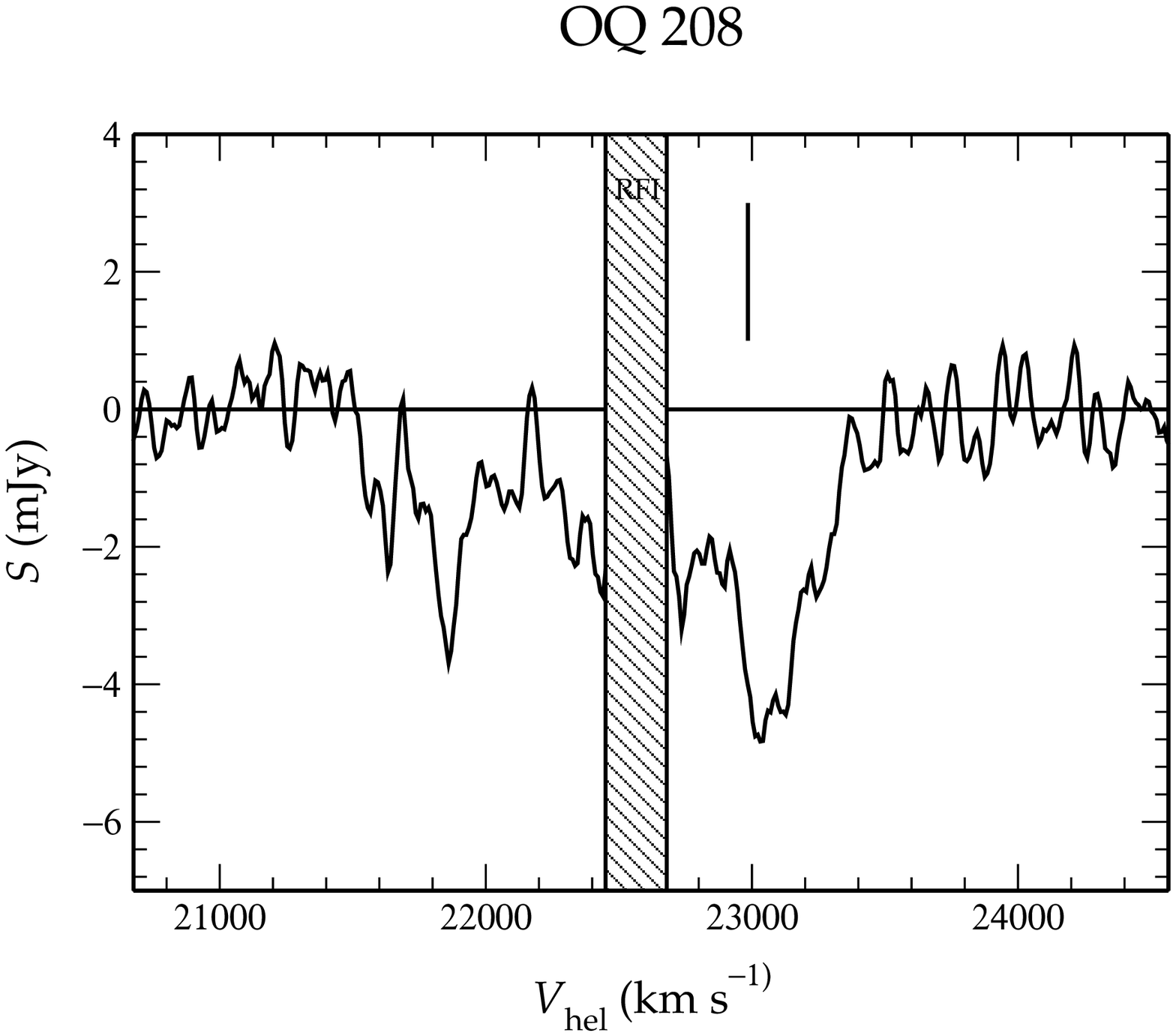,angle=0,width=4.5cm}
}
\caption{\HI\ absorption profiles detected against three radio galaxies. The observations of the radio galaxies were done using the 
WSRT, see Morganti et al. (2005) for details. The short vertical line indicates the systemic velocity.
}
\label{fig2}
\end{figure}

Other cases of fast \HI\ outflow have been found and summarised in Morganti et al. (2005), see also Fig.~\ref{fig2}.
The main result of this study is that the neutral outflows occur, in at least some cases, at kpc distance from the nucleus, and, as in the case of IC~5063,  they are most likely driven by the interactions between the expanding radio jets and the gaseous medium
enshrouding the central regions.  We estimate that the associated mass outflow rates are up to $\sim 50$ $M_\odot$ yr$^{-1}$, comparable (although at the lower end of the distribution) to the outflow rates found for starburst-driven superwinds in Ultra Luminous IR Galaxies, see Rupke et al. (2002). This suggests that massive, jet-driven outflows of neutral gas in radio-loud AGN can have as large an impact on the evolution of the host galaxies as the outflows associated with starbursts. This is important as starburst-driven winds are recognised to be responsible for inhibiting early star formation, enriching the ICM with metals and heating the ISM/IGM medium.

In the case of IC~5063, a few more parameters related to the gaseous outflow could be derived.  The mass outflow rates of cold (\HI) and warm (ionised) gas have been found to be comparable, $\sim 30$ $M_{\odot}$ yr$^{-1}$.  With a black-hole mass of $2.8 \times 10^8$ M$_\odot$, the Eddington luminosity of IC~5063 is $3.8 \times 10^{46}$ erg s$^{-1}$, this means that the kinetic power of the outflow represents only about few $\times 10^{-4}$ of the available accretion power.  This result is similar to what found for PKS~1549-79 (see below).  However, unlike the case of PKS~1549-79, IC~5063 accretes at low rate ($\dot m \sim 0.02$). Thus, in IC~5063 the kinetic power of the outflow appears to be a relative high fraction of the nuclear bolometric luminosity ($\sim 8 \times 10^{-2}$). In IC 5063, the observed outßows may have sufÞcient  kinetic power to have a signiÞcant impact on the evolution of the ISM in the host galaxy. 
 
\subsection{Young radio sources: the case of PKS~1549-79}

\begin{figure}
\centerline{\psfig{figure=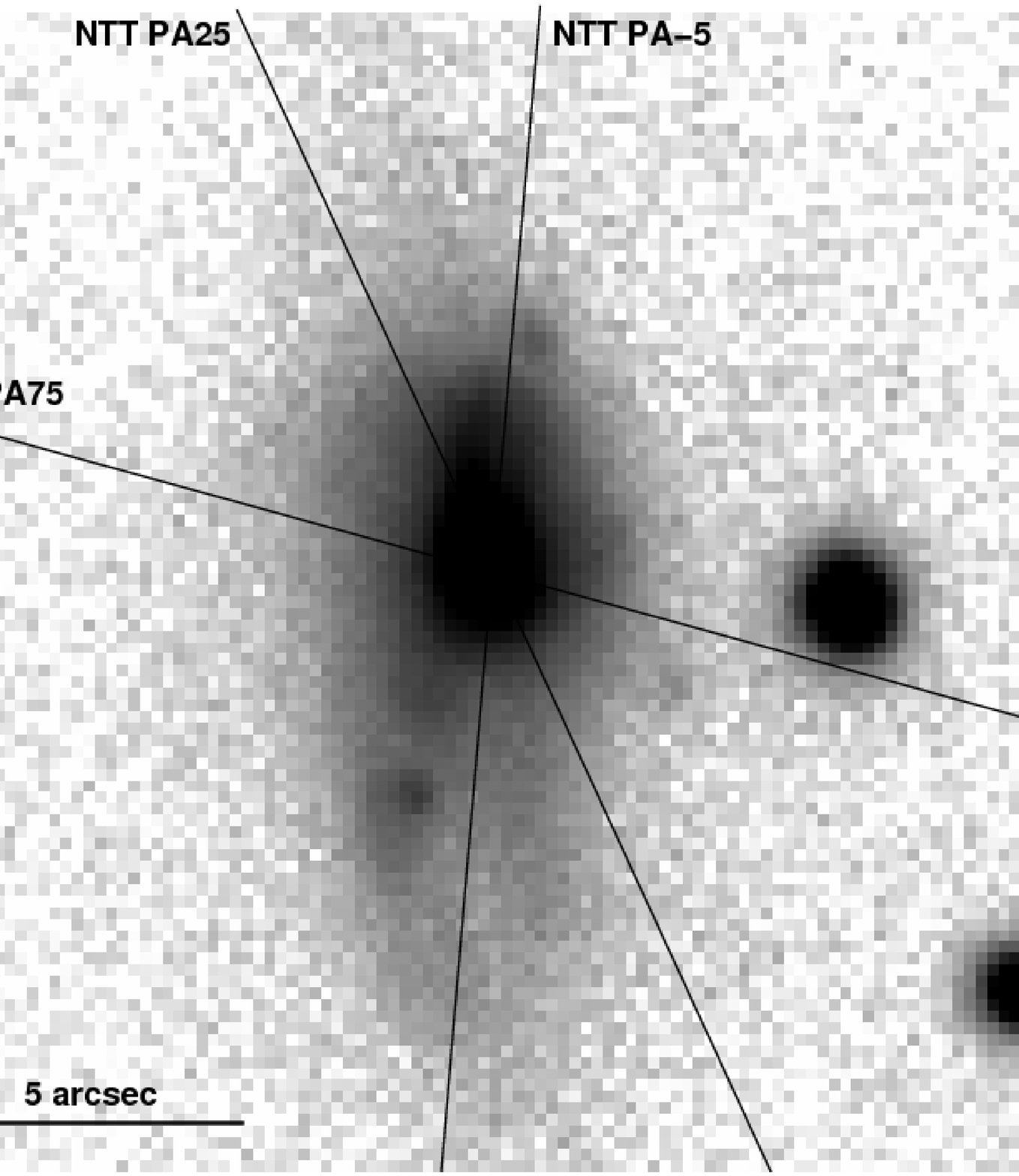,angle=0,width=4.5cm}, 
\psfig{figure=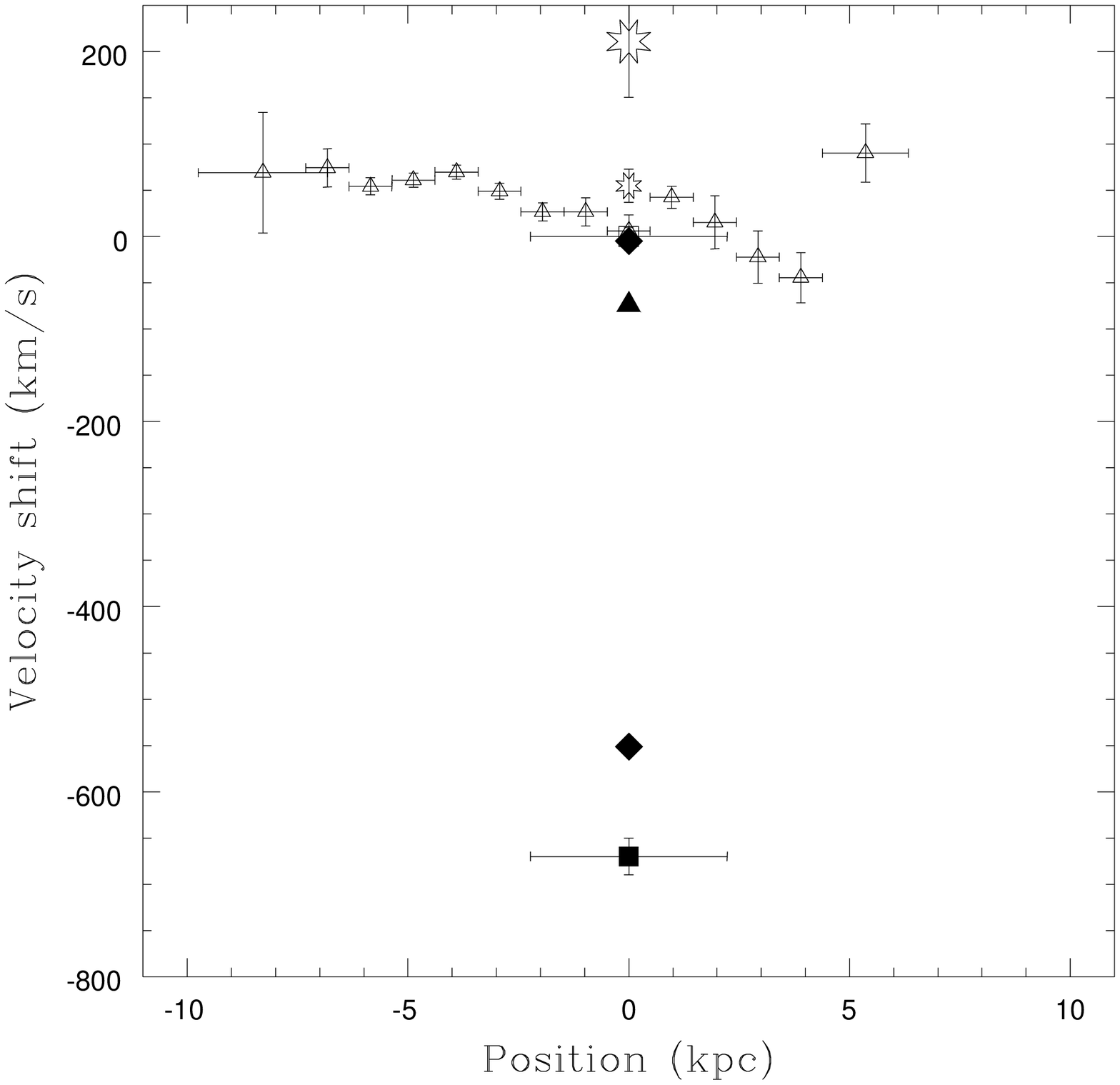,angle=0,width=4.5cm}
}
\caption{Left: Deep VLT r-image of PKS1549-79. The slit positions used for the long-slit spectroscopic observations are shown for reference. Right: Radial velocity profiles  of PKS 1549-79 obtained from the slit position PA-5. Small open  and filled triangles represent the narrow and intermediate components of H$\alpha$ respectively. 
Overplotted is the radial velocity of the deep \HI\ 21cm  absorption (large filled triangle at -30~km s$^{-1}$). More details can be found in Holt et al. (2006).
}
\label{fig3}
\end{figure}

As mentioned above, gas-rich mergers have often being identify as one of the possible trigger for AGN activity. It is not clear how common this is, but certainly in some cases  the black hole may grow rapidly through merger-induced accretion following the coalescence of the nuclei of two merging galaxies. If this is the case,  the major growth phase is largely likely to happen hidden (at optical wavelengths) by the natal gas and dust.  We have recently identified one (relatively nearby  system in such a phase of evolution  --- PKS1549-79 ($z = 0.152$).  This object shows  all the characteristics expected for a proto-quasar (see Holt et al. 2006 for details). This includes a high accretion rate onto the supermassive black hole,  a large reddening at optical wavelengths, evidence for rapid AGN-driven outflows in the  warm emission line gas and morphological evidence that the activity has been triggered in a major galaxy merger.  The signatures of this merger can be seen in tidal tails in optical (see Fig.~\ref{fig3} - left) and by the presence of a substantial young stellar population (50-250 Myr).

PKS~1549-79 is also a small ($\sim 200$ pc) and young but powerful radio galaxy.  The radio source has a core-jet structure that is considered to be quite closely aligned with the line-of-sight. Although no broad permitted (optical) lines were detected, broad Pa$\alpha$ in NIR and a reddened continuum spectrum  were observed indicating the presence of an hidden quasar nucleus this source . The presence of a fast outflow is revealed by the large blueshifts ($\Delta V \sim$680~km s$^{-1}$) and large line widths (FWHM $\sim $1300~km s$^{-1}$) of the high ionization optical emission lines (e.g. [OIII], [NeV]).  \HI\ absorption is surprisingly present indicating that it must originate not from a circumnuclear disk but from gas in which the radio source is deeply embedded. This is suggested by the fact that the VLBI observations show that the \HI\ is uniformly distributed across the radio source with nor velocity or column density  gradients. 

However, despite the evidence for rapid outflows in the warm gas , the estimated kinematic power in the warm outflow is several orders of magnitude less than required by the feedback models ($P_{kin}/L_{edd} < 10^{-4}$: Holt et al. 2006). One possible explanation for this apparent discrepancy is that much of the mass of the AGN-induced outflow is tied up in cooler or hotter phases of the interstellar medium. Unlike in objects like IC~5063, the search done so far to detect a broad absorption component of the \HI\ has not been successful but has been also been limited by the poor data quality. X-ray observations are also planned.

Interesting is the comparison of these results with what obtained in  the study of ionised gas in other young radio sources. The modelling of the complex kinematics of the gas in these sources shows that  broad components are common and they tend to be blueshifted compared to the systemic velocity of the host galaxy (derived from the extended and quiescent gas). Thus these broad components have been interpreted as gas outflows, possibly driven by jet-ISM interaction (see Holt et al. 2008 for the full discussion). Comparisons with samples in the literature also show that compact, young radio sources harbour more extreme nuclear kinematics than their extended counter-parts, a trend seen within the sample studied by Holt et al. (2008) with larger velocities in the smaller sources. 
The observed velocities are also likely to be inßuenced by source orientation with respect to the observerÕs line of sight. 
 

\section{Moving to larger scales: jet-induced star formation}

Current theoretical models  suggest that
radio source shocks propagating through the clumpy ISM/IGM trigger the collapse and/or fragmentation of overdense regions, which may then
subsequently form stars (e.g. Fragile et al 2004; Mellema, Kurk \& R\"{o}ttgering 2002, and references therein).  

Jet-induced star formation is  considered to be particularly  important for high-$z$ radio galaxies. Despite the many examples found in the nearby Universe of  gas shocked by the interaction with the radio jet, there are not many cases of jet-induced star formation known at  low-$z$. 
Nevertheless, the few cases known are ideal for study the details of such interaction. Here we discuss a few of these nearby examples.

\subsection{Jet-induced star formation in Centaurus~A}

The north-east region of Centaurus A, located about 15 kpc from its nucleus, is a particularly complex site where different structures have been found and studied. In particular, an \HI\ ring (Schiminovich et al.  1994), is situated (in projection) 
near the radio jet of Centaurus~A, as well as near very turbulent filaments of highly ionised gas and 
near regions with young stars. This is illustrated in Fig.~\ref{fig4} - left. The spatial coincidence of these  structures, together with the 
fact that the ionised and neutral gas cover the same  velocity range, has led to the suggestion (e.g.\ Graham1998) that the radio jet is interacting with the \HI\ cloud producing the turbulent gas filaments and {\sl  triggering 
the star formation in this region}.

The  higher velocity and spatial resolution of ATCA \HI\ data (Oosterloo \& Morganti, 2005) indeed reveal that, in addition to the smooth 
velocity gradient corresponding to the overall rotation of the \HI\ gas around Centaurus A, \HI\ with anomalous velocities of about 
100 \kms\  is present at the southern tip of this cloud. This is interpreted as evidence for an ongoing interaction between the 
radio jet and the \HI\ cloud. Gas stripped from the \HI\ cloud gives rise to the large filament of ionised gas and the star formation 
regions that are found downstream from the location of the interaction. The implied ßow velocities are very similar to the 
observed anomalous \HI\ velocities. Given the amount of \HI\ with anomalous kinematics and the current star formation rate, the 
efficiency of jet-induced star formation is at most of the order of a percent. 
If this  overall description is correct, the jet induced star formation is fairly inefficient. Mould et al.\ (2000) report a star formation rate for the
region of the order of a few times $10^{-3}$ \msun\ yr$^{-1}$. Assuming that the star formation rate has been constant, this implies a total mass for the
stars formed over 15 Myr (the age of the young stars) to be of order of a few times $10^4$ \msun. The amount of \HI\ showing the anomalous velocities is
about $1\times 10^6$ \msun. Thus, unless the current rate by which the \HI\ is stripped from the cloud is much higher than in the past, this appears to imply
that the efficiency of converting the  gas  stripped of the cloud into stars is at most a few percent.

\subsection{Starburst triggered by a radio jet in the Minkowsky object}

\begin{figure}
\centerline{\psfig{figure=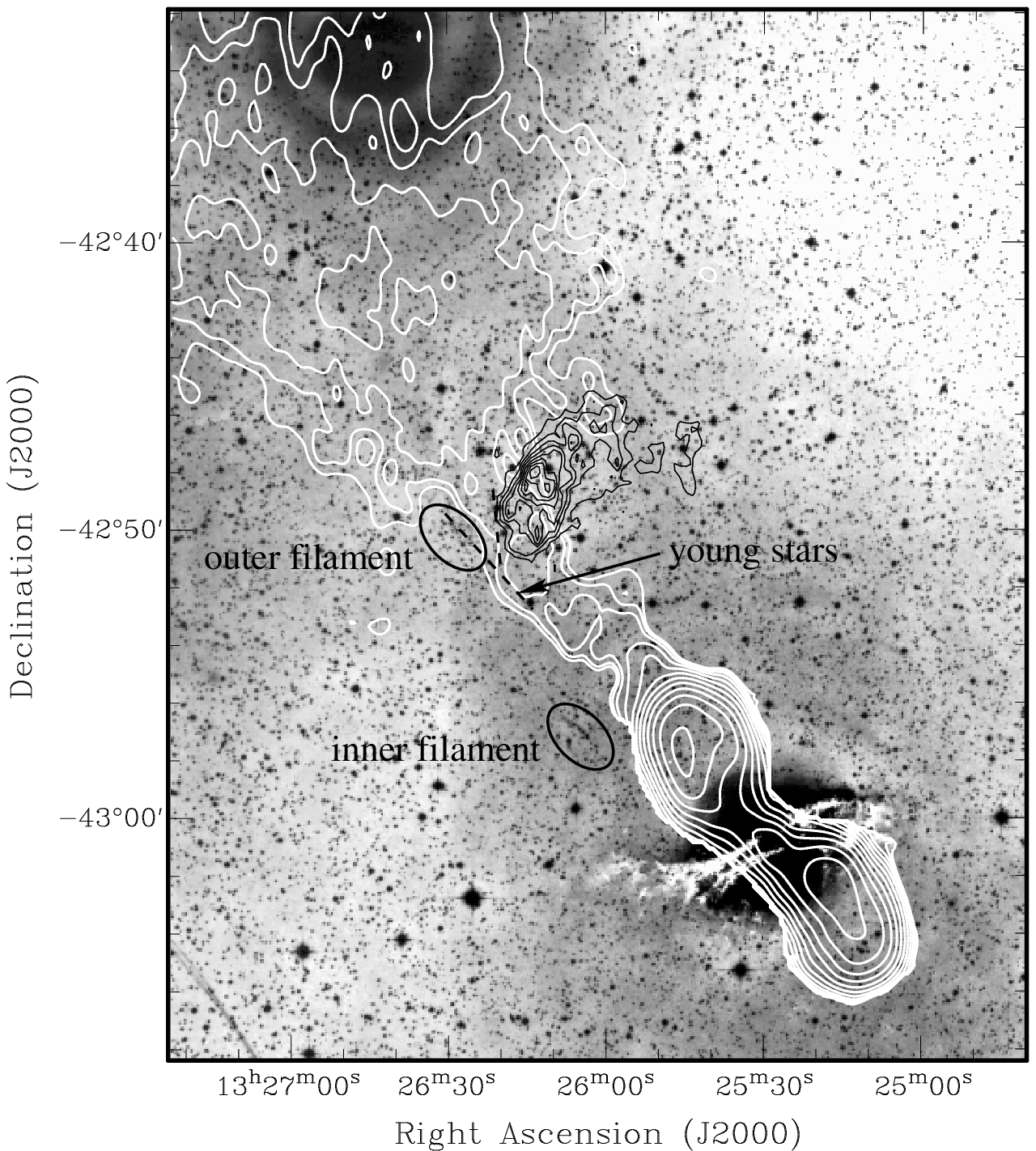,width=6cm}
\psfig{figure=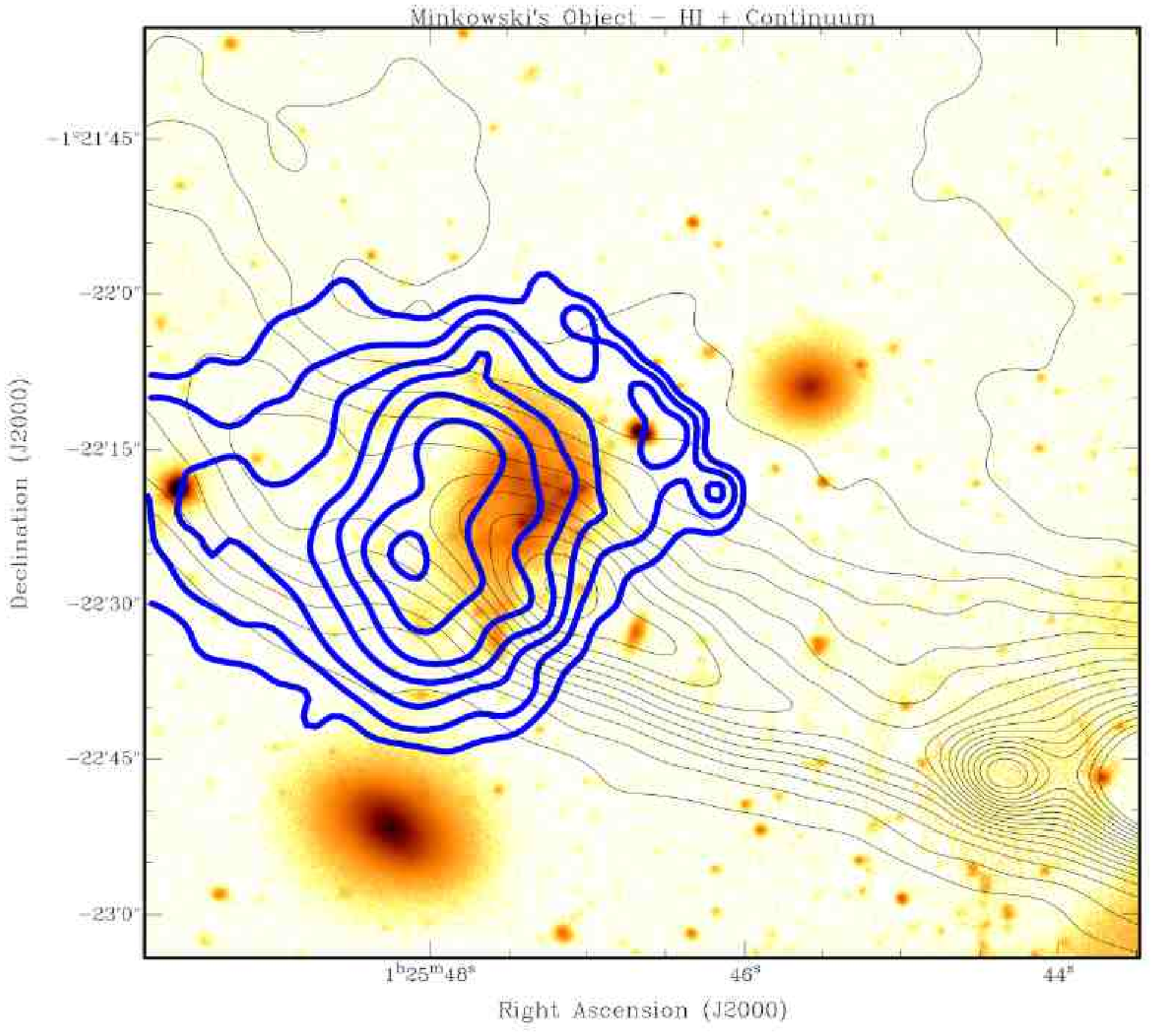,width=7cm}}
   \caption{{\sl Left:}  Overlay showing the positions of the various components of Centaurus~A.  The optical image shows the well-known dust lane of Centaurus~A and the faint diffuse optical emission that extends to very large radius. The white contours denote the radio continuum emission  The black contours denote the \HI\ cloud and the locations of the inner and outer filaments of highly ionised gas are indicated, as well as the location of young stars (from Oosterloo \& Morganti 2005).
{\sl Right:} Map of the \HI\ cloud associated with the MO, with radio continuum contours  from the VLA overlaid (from Croft et al. 2006).}
    \label{fig4}
 \end{figure}

Multi-wavebands data  have shown new evidence that the star formation in the Minkowsky object (MO), a star-forming peculiar galaxy near NGC 541, was induced by a radio jet (Crof et al. 2006). Key findings are the discovery of a 4.9x $10^8$ M$_\odot$ double \HI\  cloud straddling the radio jet downstream from MO, where the jet changes direction and de-collimates (see Fig. ~\ref{fig4} - right). 

This is similar to the jet-induced star formation associated with the Centaurus A jet, and the radio-aligned star-forming regions in powerful radio galaxies at high redshift. The age of the MO has been estimated around 7.5 Myr. While it is not possible  to completely rule out the presence of an old population in MO, the data are consistent with MO having formed de novo when the jet interacted with the ambient ISM/IGM.  Unlike Centaurus A, we propose that the \HI\ associated with MO formed in situ through cooling of clumpy, warm gas, in the stellar bridge or cluster IGM, as it was compressed by  radiative shocks at the jet collision site, in agreement with numerical simulations (Fragile et al. 2004). The star formation in MO then followed from the cooling and collapse of such \HI\  clouds, and the \HI\  kinematics, which show 40 \kms\ shear velocities, are also consistent with such models.

\subsection{A new candidate:  PKS~2250-41}

\begin{figure}
\centerline{\psfig{figure=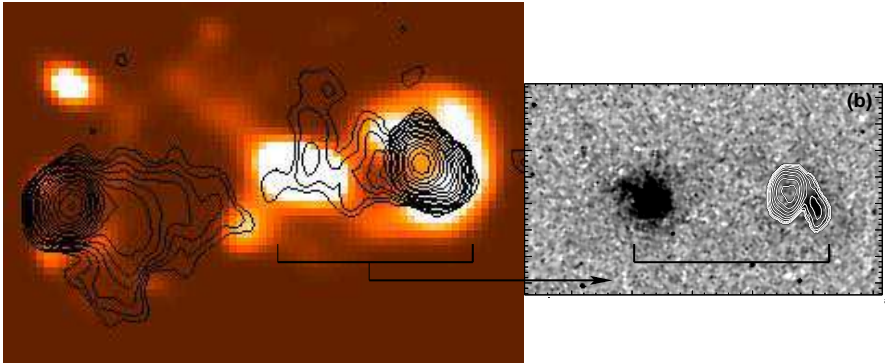,width=12cm}}
   \caption{{\sl Left:} Low resolution ATCA radio image superimposed to narrow band image (Clark et al. 1997); {\sl Right:} high resolution WFPC2 F547M continuum image of PKS~2250-41 overlaid  with 15~GHz VLA radio contours (from Inskip et al. 2008).
The  optical (as well as the infrared) image displays ßux at the position  of peak emission line intensity in the western arc, which is also 
coincident with the secondary hotspot within the western radio lobe.}
    \label{fig5}
 \end{figure}

PKS~2250-41 is an archetypal example of a galaxy displaying jet-cloud interactions, with clear evidence for shocks associated with the
expanding radio source. This is illustrated by the observed distribution of ionised gas (see  fig.~\ref{fig5}-left) and by the variation in ionisation state and gas kinematics in the vicinity of the western radio lobe hotspots.     
Past studies of PKS2250-41 have suggested that the prominent emission line arc to the west of the host galaxy originates from a direct collision
between the radio source jet and a companion galaxy (Clark et al 1997, Villar-Martin et al. 1999). 
The primary evidence is the continuum emission which, in addition to the well-studied line emission, is also observed within the western
radio lobe.  The continuum emission is approximately co-spatial with the secondary radio source hotspot (Fig.~\ref{fig5}-right), has low polarization and only a limited contribution of nebular continuum emission.  Clark et al (1997) suggested that the residual continuum emission could originate from a late-type spiral or irregular galaxy, with which the radio jet has collided; radio source shocks driven through the gas clouds associated with such an object can also account for the impressive scale and luminosity of the observed emission line arc, and possibly also the shortness of the western lobe relative to the eastern lobe. 
New optical and infrared observations of PKS~2250-41 add further weight to this scenario (see Inskip et al. 2007 for all details). 
Continuum emission is detected in both the $K_S$ and on the WFPC2 F547M filters within the arc, coincident with the
secondary radio hotspot. Figure~\ref{fig5}-right  displays the  high resolution optical
continuum  overlaid with radio contours.     However, and interestingly, the inferred spectral shape of the continuum  implies that a {\it very} young stellar population is dominating the optical
emission. The age of the  unreddened stellar populations has been estimated  between either 0.006-0.009 Gyr or
0.05-0.1 Gyr, or alternatively reddened  young stellar  population (YSP) of even younger ages.
This  suggests that the proximity of the radio source may very well have triggered recent star formation within this object.  

\section{Conclusions}

Clear evidence of the impact of the interaction between the  radio jets and the ISM have been found from the study of the kinematics of the gas in radio-loud sources. 
Fast outflows of neutral atomic hydrogen and ionised gas are produced by such interaction. The presence of neutral gas associated with such outflows indicates that the gas can cool very efficiently following a strong jet-cloud interaction.
Outflows of similar velocities are observed in \HI\ and in ionised gas but the mass outflow rate is relatively high in \HI\ and much less in ionised gas.
The derived mass outflow rate in \HI\ ranges between a few and $\sim 50$ M$_\odot$/yr  comparable (although at the lower end) to that found in Ultraluminous IR galaxies. Thus, jet-driven outflows can have a similar impact on the evolution of a galaxy as starburst-driven superwinds.
However, the importance for the feedback is not completely clear. In IC~5063 the outflow energy is a reasonable fraction of the nuclear bolometric luminosity (but not of the Eddington luminosity). On the other hand, PKS 1549-79 is in a stage where the nucleus is still hidden (in the optical) by the gas/dust coming from the merger that triggered the radio source. However,  the outflow of ionised gas is  not as large as expected in the quasars feedback model (while \HI\ outflow has not been found yet in this radio source). Thus, this study  so far indicates that outflows of ionised gas are typically not massive enough to clear the nuclear gas in young radio sources while the situation is more promising for outflows of cold gas although at the moment the statistics about occurrence of such outflows is limited. This result appears to be confirmed also for other objects (see also Tadhunter (2008) for an overview). 

Jet-induced star formation have been found in a very limited number of nearby radio galaxies. We have presented here a possible new case. 
The jet induced star formation appears to be relatively inefficient both in the case of Cen A and in the MO. Comparing the global star formation efÞciency  $M_{\rm stars}/M_{\rm \HI}$ in MO  we Þnd  $\sim 4$\%,  which is similar to that in Centaurus A (a few percent; Oosterloo 
\& Morganti 2005).  However, different origins have been suggested for the two systems.  In situ formation through cooling of clumpy, warm gas has been suggested in the case of Minkowsky object, while in the case of Cen~A the \HI\ was likely already present as large scale structure. Unfortunately, no \HI\ information is available for PKS~2250-41 because the system is at too high redshift. 

In summary, \HI\ and ionised gas observations of radio-loud sources provide extra constraints on the effects of this kind of
AGN on the ISM. Although we are still building up the statistics for a large number of objects, these studies are particularly 
important in order to get a more complete and realistic picture of the effects of feedback in galaxy evolution.

\bigskip

{\sl The author would like to thank the organisers of this very interesting meeting for the kind invitation. The author  would also like to acknowledge and thank the main  collaborators involved in the projects presented here:  Clive Tadhunter, Tom Oosterloo, Joanna Holt, Katherine Inskip, Steve Croft and Wil van Breugel.}


\end{document}